\documentclass[24pt]{article}
\usepackage{amssymb}
\usepackage{amsfonts}
\usepackage{amssymb,amsmath}
\usepackage{mathrsfs}
\usepackage{latexsym}
\usepackage{amsmath}
\usepackage{latexsym}
\usepackage[cp1251]{inputenc}
\usepackage{graphicx}
\usepackage{color}

\title{Application of a quantum wave impedance method for zero-range singular potentials}
\author{O. I. Hryhorchak\\
{\small Department for Theoretical Physics, Ivan Franko National
University of Lviv,}\\
{\small 12, Drahomanov Str., Lviv, UA--79005,
Ukraine}\\
\small{\it{Orest.Hryhorchak@lnu.edu.ua}}}

\def\th{\mathop{\rm th}\nolimits}
\def\sign{\mathop{\rm sign}\nolimits}
\def\cth{\mathop{\rm cth}\nolimits}

\def\ctg{\mathop{\rm ctg}\nolimits}

\begin{document}
\renewcommand{\abstractname}{Abstract}
\maketitle

\begin{abstract}
An application of a quantum wave impedance method for a study of quantum-mechanical systems which con\-tain singular zero-range potentials is considered. It was shown how to reformulate the problem of an investigation of mentioned systems in terms of a quantum wave impedance. As a result both the scattering and bound states problems are solved for systems of single $\delta$, double $\delta$ and single $\delta-\delta'$ potentials. The formalization of solving systems with an arbitrary combination of a piesewise constant potential and a $\delta$-potentials with the help of a quantum wave impedance approach is described. 
\end{abstract}

\section{Introduction}

Singular zero-range potentials are widely used for modelling of real quan\-tum mechanical systems and are
widely discussed in a literature \cite{Demkov_Ostrovskii:1988, Albeverio_etall:2005}. One of the most famous examples is a Dirac comb \cite{Cordoba:1989, Sfiat:2013} which is exactly solvable model and shows a mechanism of a zone spectra formation. After a pioneering paper of Berezin and Faddeev \cite{Berezin_Faddeev:1961} a lot of articles on this topic have appeared. It was shown how singular zero-range potentials, in particular a $\delta$-potential, can be applied for a modelling of different physical systems, namely quantum waveguides \cite{Albeverio_Cacciapuoti_Finco:2007, Cacciapuoti_Exner:2007}, spectral filters \cite{Turek_Cheon:2012, Turek_Cheon:2013}, infinitesimally thin sheets \cite{Zolotaryuk_Zolotaryuk:2015, Zolotaryuk:2013}, an entanglement of polymers \cite{Ferrari_Vakhtang_Vakhtang:2005}, propagation of a light  \cite{Zurita_Halevi:2000, Lin_Jao:2006}, a Bose-Einstein condensation in a harmonic trap \cite{Uncu_etall:2007},
supersymmetry \cite{Diaz_etall:1999} as well as  systems within area of  condensed matter physics \cite{Cheon_Exner_Seba:2000} and \mbox{others \cite{Kostenko_Malamud:2013}.}

An idea to enrich a $\delta$-potential with its first derivative appeared quite a long time ago. And we should recognize that it
extended the scope of physical systems which can be modelled by the help of this new singular zero-range potential. For example, a $\delta-\delta'$-potential is used effectively to model multilayer structures, in particular a typical transistor in a zero-thickness limit which in this case called a ``point'' transistor \cite{Zolotaryuk_Zolotaryuk:2015, Zolotaryuk_Tsironis_Zolotaryuk:2019}. In this $\delta-\delta'$ model there are several extremely sharp peaks (when a transmission probability is not zero) at specific values of an applied voltage while beyond these values of an applied voltage the system is almost completely reflective. These peaks depend not only on an applied voltage but also on parameters of the model.

But it is also well-known that the interpretation of a $\delta'$-potential is not clear and there were a lot of discussions about what $\delta'$-potential really is? The similar situation is with a $\delta$-potential in 2D and higher dimensions. So while we deal with a $\delta$-potential in 1D we have no problems with its definition but in a case of 2D or 3D it is necessary to make an appropriate regularization.

For a $\delta'$-potential there are at least four different approaches for its definition. In many articles \cite{Albeverio_etall:1984, ALBEVERIO_DABROWSKI_KURASOV:1993, ALBEVERIO_DABROWSKI_KURASOV:1998, Albeveriot_Brzeiniakts_Darowski:1994, Coutinho_Nogami_Perez:1997, Coutinho_etall:1999, Coutinho_Nogami_Tomio:1999}
it is considered only as a label for specific boundary conditions for a wave function or its first derivative, in others \cite{Coutinho_Nogami_Toyama:2009, Patil:1994} it is interpreted as a dipole interaction or as a rectangular approximation \cite{Christiansen_etall:2012, Christiansen:2003, Zolotaryuk_Christiansen_Iermakova:2006}. In papers \cite{Golovaty2012, Golovaty_Hryniv:2012} it is defined as a zero limit of some smooth potential.
Each of them has its drawbacks and relates to different physical realities. An analysis of these you can find in a paper \cite{Lange:2015}.

In this article we are not going to go deeper in mathematical aspects of a $\delta'$-potential.
There are a lot of articles which can help one to become familiar with the topic of zero-range potentials \cite{Zolotaryuk_Zolotaryuk:2011, Zolotaryuk:2017, Zolotaryuk_Christiansen_Iermakova:2007, Zolotaryuk:2010, MunozCastaneda_Nietoab_Romaniega:2019, Gadella_Glasser_Nieto:2011, Nyeo:2000, Coutinho_Nogami_Perez:1997, Jacobs:2016, Jacobs:2019, Lange:2015, Gadella:2009, Gadella:2011, Shoemaker:1972, Toyama:2007, Lee_etall:2016, Yong_Son:2018, Cacciapuoti_Mantile_Posilicano:2016, Vincenzo_Sanchez:2010_1, Vincenzo_Sanchez:2016, Fatih_Gadella_Uncu:2020}.
It is worth to say that different meanings of a $\delta'$-potential bear different regularization procedures and
different physically observable values, namely different values of reflection and transmission coefficients \cite{Christiansen_etall:2003,Toyama:2007,Christiansen_etall:2012, Ahmed_etall:2016}.

The problems with a definition of a $\delta'$-potential were mathematically clarified by Kurasov and co-authors \cite{Kurasov:1996, ALBEVERIO_DABROWSKI_KURASOV:1998}. Their approach gives the relation
between singular potentials (namely,  a $\delta'$ potential) and matching conditions at the origin.

Besides $\delta$ and $\delta'$ potentials one can face also with a $\delta''$-potential in the literature \cite{Patil:1994, Zolotaryuk:2015}. But we are not going to consider it here because generally it gives nothing significantly new for us.
	
Our aim is to reformulate the problem of an investigation of quantum mechanical systems with zero-range potentials in terms of a quantum wave impedance method and to show how significantly the solving of such systems can be simplified compared to the classical approaches.   

\section{Matching condition for a quantum wave\\ impedance in a case of $\delta$-potential}
In model systems the most common singular zero-range potential is a $\delta$-potential. It can be defined in different ways as a marginal case of a non-singular potential. One of them is  as follows
\begin{eqnarray}
\delta(x-a)=\lim_{\varepsilon\rightarrow 0}\frac{1}{\varepsilon\sqrt{\pi}}\exp\left[-(x-a)^2/\varepsilon^2\right].
\end{eqnarray} 
But $\delta$-potential significantly differ from non-singular potentials, namely, a $\delta$-potential causes a break of a first derivative of a wave-function at a point where it is located. It means that we have a break of a quantum wave impedance function at the origin of a $\delta$-potential. To get a matching condition for a quantum wave impedance at this point we integrate both sides of the equation for a quantum wave impedance \cite{Arx1:2020} with a potential $U(x)=\alpha\delta(x-a)$:
\begin{eqnarray}
\lim_{\varepsilon\rightarrow 0}\int\limits_{-\varepsilon}^{\varepsilon}\left(\frac{dZ(x)}{dx}+i\frac{m}{\hbar}Z^2(x)\right)=\lim_{\varepsilon\rightarrow 0}\int\limits_{-\varepsilon}^{\varepsilon}i\frac{2}{\hbar}\left(E-\alpha\delta(x-a)\right).
\end{eqnarray}
It gives
\begin{eqnarray}\label{Z_delta_matching_cond}
Z(a+0)-Z(a-0)=-\frac{2i\alpha}{\hbar},
\end{eqnarray}
where $Z(a+0)$ means the right-side boundary of a quantum wave impedance function at a point $a$ and $Z(a-0)$ is the left-side one. We will widely use this condition in the next sections.

\section{Bound states case for a single $\delta$-well}
In this section and in a next one we consider how a concept of a quantum wave impedance can be applied to both a bound states case and a scattering case of a single $\delta$-potential.

In a bound states case the equation for a quantum wave impedance can be written as follows
\begin{eqnarray}\label{Zdelta}
\frac{dZ(x)}{dx}+i\frac{m}{\hbar}Z^2(x)=i\frac{2}{\hbar}
\left(\frac{}{}E+\alpha\delta(x)\right).
\end{eqnarray}
We find a solution in a form: $Z(x)=C\sign(x)$, where
\begin{eqnarray}
\sign(x)=\left\{\begin{array}{c}
1, \ \ x\geq 0\\
-1, \ \  x<0
\end{array}\right. .
\end{eqnarray}
Taking into account that 
\begin{eqnarray}
\frac{d \sign(x)}{dx}=2\delta(x)
\end{eqnarray}
and substituting it in a formula (\ref{Zdelta}) we get that 
\begin{eqnarray}
C=\frac{i\alpha}{\hbar}, \qquad E=-\frac{m\alpha^2}{2\hbar^2}.
\end{eqnarray}
Thus, we have
\begin{eqnarray}
Z(x)=\frac{i\alpha}{\hbar}\sign(x).
\end{eqnarray}
In the $x\rightarrow\pm\infty$ limits we have correct results for $Z(x)$ according to \cite{Arx1:2020}.

Because $E<0$ it is a bound states case. We can find its wave function on the base of a following relation
\begin{eqnarray}
\frac{\hbar}{im}\frac{\psi'(x)}{\psi(x)}=\frac{i\alpha}{\hbar} \sign(x).
\end{eqnarray}
Integrating both sides of this relation we get
\begin{eqnarray}
\ln[\psi(x)]= \pm\frac{ m\alpha}{\hbar^2}|x|+\ln(A),
\end{eqnarray}
or finally
\begin{eqnarray}
\psi(x)=A\exp\left[-\frac{ m\alpha}{\hbar^2}|x|\right].
\end{eqnarray}
Constant $A$ can be found from the normalization condition \mbox{$\int\limits_{-\infty}^\infty|\psi(x)|^2=1$}. 
Thus, we have
\begin{eqnarray}
\psi(x)=\frac{\sqrt{m\alpha}}{\hbar}\exp\left[\pm \frac{ m\alpha}{\hbar^2}|x|\right].
\end{eqnarray}

\section{Scattering case for a single $\delta$-barrier}

For a $\delta$-barrier the equation for  a quantum wave impedance is as follows:
\begin{eqnarray}
\frac{dZ(x)}{dx}+i\frac{m}{\hbar}Z^2(x)=i\frac{2}{\hbar}(E+\alpha\delta(x)).
\end{eqnarray}
Assuming that the wave incidents on the left of a $\delta$-barrier the solution of this equation is
\begin{eqnarray}
Z(x)=\left\{\begin{array}{cc}
z_0\th\left[\gamma_0 x+\phi\right], & x<0\\
z_0, &  x>0
\end{array}\right..
\end{eqnarray}
Using the matching condition for a $\delta$-potential (\ref{Z_delta_matching_cond}) we find
\begin{eqnarray}
z_0-z_0\th[\phi]=-i\frac{2\alpha}{\hbar},\qquad
\th[\phi]=1+i\frac{2\alpha}{z_0\hbar}.
\end{eqnarray}
Taking into account that
\begin{eqnarray}
\th\left[\gamma_0 x+\phi\right]=\frac{\th[\gamma_0 x]+\th[\phi]}{1+\th[\gamma_0 x]\th[\phi]}
\end{eqnarray}
and
\begin{eqnarray}
\th[\phi]=\frac{1-\exp[-2\phi]}{1+\exp[-2\phi]}
\end{eqnarray}
we get
\begin{eqnarray}
Z(x)=z_0\frac{\th[\gamma_0 x]+1+i\frac{2\alpha}{z_0\hbar}}{1+\th[\gamma_0 x]\left(1+i\frac{2\alpha}{z_0\hbar}\right)}, 
\quad x<0
\end{eqnarray}
and a wave reflection amplitude coefficient 
\begin{eqnarray}
r=\exp[-2\phi]=-\frac{i\alpha}{z_0\hbar+i\alpha}=\frac{-i\alpha m}{i\alpha m+\hbar^2k_0},
\end{eqnarray}
where $k_0=mz_0/\hbar=\sqrt{2mE}/\hbar$ is a wave vector.
A wave transmission amplitude coefficient $t$ is equal to
$r+1$ because the thickness of a $\delta$-barrier is equal to zero. Thus
\begin{eqnarray}
t=\frac{z_0\hbar}{z_0\hbar+i\alpha}=\frac{\hbar^2k_0}{\hbar^2k_0+i\alpha m}.
\end{eqnarray}
It also gives a reflection $R$ and a transmission $T$ coefficients:
\begin{eqnarray}
R=|r|^2=\frac{\alpha^2}{z_0^2\hbar^2+\alpha^2}=
\frac{1}{1+\frac{2\hbar^2E}{m\alpha^2}}
,\nonumber\\
T=|t|^2=\frac{z_0^2 \hbar^2}{z_0^2\hbar^2+\alpha^2}
=
\frac{1}{1+\frac{m\alpha^2}{2\hbar^2E}}.
\end{eqnarray}
We see that $R$+$T$=1.

\section{Scattering case for a double $\delta$-barrier po\-tential}
Consider a model which is a little more complicated than the previous one, namely, the model of a double $\delta$-barrier:
\begin{eqnarray}
U(x)=\alpha(\delta(x-a)+\delta(x+a)), \:\alpha>0.
\end{eqnarray}
In this case a solution for a quantum wave impedance  is
\begin{eqnarray}
Z(x)=\left\{\begin{array}{cc}
z_0\th\left[\gamma_0 x+\phi_1\right], & x<0\\
z_0\th\left[\gamma_0 x+\phi_2\right], & 0<x<a\\
z_0, &  x>0
\end{array}\right..
\end{eqnarray}
Thus, we have
\begin{eqnarray}
\th\left[\phi_2\right]-\th\left[\phi_1\right]=
-i\frac{2\alpha}{z_0\hbar},\nonumber\\
1-\th\left[\gamma_0 a+\phi_2\right]=-i\frac{2\alpha}{z_0\hbar}.
\end{eqnarray}
It gives 
\begin{eqnarray}
\th[\phi_1]=
\frac{\left(-i\frac{2\alpha}{z_0\hbar}\right)^2+i\frac{2\alpha}{z_0\hbar}-1+\cth[\gamma_0 a]}{-i\frac{2\alpha}{z_0\hbar}-1+\cth[\gamma_0 a]},
\end{eqnarray}
\begin{eqnarray}
r=\exp(-2\phi_1)=-\frac{-i\frac{2\alpha}{z_0\hbar}\left(-i\frac{2\alpha}{z_0\hbar}-2\right)}{\left(-i\frac{2\alpha}{z_0\hbar}\right)^2-2+\cth[\gamma_0 a]}.
\end{eqnarray}
Reminding that $\gamma_0$ is imaginary and introducing $k_0=\frac{mz_0}{\hbar}=\gamma_0/i$ ($k_0$ is real) we get 
\begin{eqnarray}
R=
\frac{16G(G+1)}{4(G+1)^2+\ctg[k_0a]},\qquad
T=\frac{4+\ctg[k_0a]}{4(G+1)^2+\ctg[k_0a]},
\end{eqnarray}
where
\begin{eqnarray}
G=\frac{\alpha^2m^2}{k_0^2\hbar^4}=\frac{\alpha^2m}{2E\hbar^2}.
\end{eqnarray}

\section{Bound states case for a double $\delta$-well \mbox{potential}}
In this section we consider a model of a double $\delta$-well:
\begin{eqnarray}
U(x)=-\alpha(\delta(x-a)+\delta(x+a)), \:\alpha>0.
\end{eqnarray}
Our task is to find energies and wave functions of bound states of this system.
In a region $-a<x<a$ the solution of the equation for a quantum wave impedance function \cite{Arx1:2020, Arx2:2020} is
\begin{eqnarray}
Z(x)=z_0\th[\varkappa_0x+\phi],
\end{eqnarray}
where $z_0=\sqrt{2E/m}$ is the characteristic impedance, $\varkappa_0=\frac{m}{i\hbar}z_0=\sqrt{-2mE}/\hbar>0$ (it is because $E<0$ and $z_0$ is imaginary) and $\phi$ is the phase to be calculated. To the left of a point $x=-a$ we have $Z(x)=-z_0, x<-a$ and to the right of a point $x=a$ a quantum wave impedance is $Z(x)=z_0, x>a$. So taking into account the matching conditions (\ref{Z_delta_matching_cond}) we get
\begin{eqnarray}
&&z_0\th[-\varkappa_0a+\phi]+z_0=i\frac{2\alpha}{\hbar},\nonumber\\
&&z_0-\th[\varkappa_0a+\phi]=i\frac{2\alpha}{\hbar}.
\end{eqnarray}
Reminding that $z_0=\hbar \varkappa_0/im$ and after simple transformations the previous system of equations will take the following form: 
\begin{eqnarray}
-\frac{\th[\varkappa_0a]-\th[\phi]}{1-\th[\varkappa_0a]\th[\phi]}+1=\frac{2m\alpha}{\hbar^2\varkappa_0},\nonumber\\
1-\frac{\th[\varkappa_0a]+\th[\phi]}{1+\th[\varkappa_0a]\th[\phi]}=\frac{2m\alpha}{\hbar^2\varkappa_0}.
\end{eqnarray}
We have two different solutions of this system of equations:
\begin{eqnarray}
\phi=0,\quad 1-\th[\varkappa_0a]=\frac{2m\alpha}{\hbar^2\varkappa_0},\nonumber\\
\phi=i\frac{\pi}{2},\quad 1-\cth[\varkappa_0a]=\frac{2m\alpha}{\hbar^2\varkappa_0}.
\end{eqnarray}
It gives well-known relations for an energy of two bound states in the system of a double $\delta$-well:
\begin{eqnarray}
\frac{\hbar^2\varkappa_0}{m\alpha}=1+\exp[-2\varkappa_0a],\quad
\frac{\hbar^2\varkappa_0}{m\alpha}=1-\exp[-2\varkappa_0a].
\end{eqnarray}
These two energy values of bound states correspond to two different quantum wave impedance functions, namely: 
\begin{eqnarray}
Z_1(x)=\left\{\begin{array}{cc}
-iz_0=-i\sqrt{2E/m}, & x< a\\
z_0\th[\varkappa_0x], &  -a<x<a \\
iz_0=i\sqrt{2E/m},& x>a
\end{array}\right.,
\end{eqnarray}
\begin{eqnarray}
Z_2(x)=\left\{\begin{array}{cc}
-iz_0=-i\sqrt{2E/m}, & x< a\\
z_0\cth[\varkappa_0x],&  -a<x<a \\
iz_0=i\sqrt{2E/m},& x>a
\end{array}\right..
\end{eqnarray}
Using formula which relates a quantum wave impedance function and a wave function \cite{Arx1:2020} on the base of expressions for $Z_1(x)$ and $Z_2(x)$ we get two wave-functions of the bound states:
\begin{eqnarray}
\psi_1(x)&=&\exp\left[\frac{im}{\hbar}\int\left(\frac{}{}\!\!\!-z_0\theta(-x-a)+z_0\theta(x-a)+\right.\right.\left.\left. z_0\th(\varkappa_0 x)\left(
\theta(x+a)-\theta(x-a)\right)\frac{}{}\right)dx \right]=\nonumber\\
&=&A_1\left(\frac{}{}\exp(-\varkappa_0|x-a|)+\exp(-\varkappa_0|x+a|)\right)
\end{eqnarray}
and
\begin{eqnarray}
\psi_2(x)&=&\exp\left[\frac{im}{\hbar}\int\left(\frac{}{}\!\!\!-z_0\theta(-x-a)+z_0\theta(x-a)+\right.\right.\left.\left. z_0\cth(\varkappa_0 x)\left(
\theta(x+a)-\theta(x-a)\right)\frac{}{}\right)dx \right]=\nonumber\\
&=&A_2\left(\frac{}{}\exp(-\varkappa_0|x-a|)-\exp(-\varkappa_0|x+a|)\right).
\end{eqnarray}
Constants $A_1$ and $A_2$ can be obtained from the normalization condition: $\int\limits_{-\infty}^\infty|\psi_1|dx=\int\limits_{-\infty}^\infty|\psi_2|dx=1$.
The same approach we can use to solve the bound states problem for a system of three or more $\delta$-wells.

\section{Matching condition for a quantum \\ wave impedance in a case of a $\delta\!-\!\delta'$-potential}

In this and next two chapters we deal with the model system with a potential energy of the following form:
\begin{eqnarray}
U(x)=-\alpha\delta(x-a)+\beta\delta'(x-a),
\end{eqnarray}
which contains both $\delta$ and $\delta'$ functions with arbitrary constants $\alpha>0$ \mbox{and $\beta$}.
Notice that a zero-range singular $\delta'$-potential
causes not only a break of a first derivative of a wave function like a $\delta$-potential but also a break of a wave function itself. Thus, there are a lot of discussions about its applicability for modelling of real physical systems mainly because the values of physical quantities calculated in the models with a $\delta'$-potential depend on a regularization procedure. It is not our aim to clarify this topic here. 
Here we will follow the approach of Kurasov and co-authors \cite{Kurasov:1996, ALBEVERIO_DABROWSKI_KURASOV:1998} who proposed to consider only such wave-functions $\psi(x)$ which belong to the subspace of
a self-adjoint extension of a kinetic energy operator: $-\frac{\hbar^2}{2m}\frac{d^2}{dx^2}$. In this case the mentioned problem of choosing the regularization procedure fades out. These functions $\psi(x)$, as it was well-established earlier (for example here \cite{Gadella:2009}) have to satisfy the following matching conditions at the origin:
\begin{eqnarray}\label{psi_d_d'_mc}
\begin{pmatrix}
\psi(a+0) 
\\
\psi'(a+0) 
\end{pmatrix}
=
\begin{pmatrix}
\frac{1+\tilde{\beta}}{1-\tilde{\beta}}&  0 
\\
-\frac{\tilde{\alpha}}{1-\tilde{\beta}^2}& 
\frac{1-\tilde{\beta}}{1+\tilde{\beta}} 
\end{pmatrix}
\begin{pmatrix}
\psi(a-0) 
\\
\psi'(a-0) 
\end{pmatrix},
\end{eqnarray}
where $\psi(a+0)$, $\psi(a-0)$, $\psi'(a+0)$, $\psi'(a-0)$ are right-side and left-side limits of a function $\psi(x)$ and its first derivative at point $x=a$. $\tilde{\alpha}$ and $\tilde{\beta}$ are the notations for 
\begin{eqnarray}
\tilde{\beta}=\frac{m\beta}{\hbar^2},\qquad \tilde{\alpha}=\frac{2m\alpha}{\hbar^2}.
\end{eqnarray}
Matching conditions (\ref{psi_d_d'_mc}) also define the rules of $\delta$ and $\delta'$ action on a wave function:
\begin{eqnarray}
\psi(x)\delta(x-a)&=&\psi(a)\delta(x-a),\nonumber\\
\psi(x)\delta'(x-a)&=&\psi(a)\delta'(x-a)-\psi'(a)\delta(x-a).
\end{eqnarray}
On the base of matching conditions (\ref{psi_d_d'_mc}) for a wave function and its first derivative we get one matching condition for a quantum wave impedance function at the origin of a $-\alpha\delta(x-a)+\beta\delta'(x)$ potential
\begin{eqnarray}\label{Z_d_d'_mc}
Z(a+0)=\frac{i\hbar\tilde{\alpha}}{m(1+\tilde{\beta})^2}+
\frac{(1-\tilde{\beta})^2}
{(1+\tilde{\beta})^2}Z(a-0).
\end{eqnarray}

\section{Bound states case for a $\delta-\delta'$ potential}

The solution of the equation for a quantum wave impedance with a $\delta-\delta'$ potential
\begin{eqnarray}
\frac{dZ(x)}{dx}+i\frac{m}{\hbar}Z^2(x)=i\frac{2}{\hbar}(E+\alpha\delta(x)-\beta\delta'(x)).
\end{eqnarray}
in a bound states case ($E<0$) is as follows
\begin{eqnarray}
Z(x)=\left\{\begin{array}{cc}
z_0=i\sqrt{2|E|/m}, & x>0\\
-z_0=-i\sqrt{2|E|/m}, &  x<0
\end{array}\right..
\end{eqnarray} 
Using the matching condition (\ref{Z_d_d'_mc}) we get the relation for an energy of a bound state
\begin{eqnarray}
z_0=\frac{i\hbar\tilde{\alpha}}{m(1+\tilde{\beta})^2}-
\frac{(1-\tilde{\beta})^2}
{(1+\tilde{\beta})^2}z_0
\end{eqnarray}
or
\begin{eqnarray}
z_0=\frac{i\hbar\tilde{\alpha}}{2m(1+\tilde{\beta}^2)},
\end{eqnarray}
which for an energy $E$ of a bound state gives
\begin{eqnarray}
E=-\frac{\hbar^2\tilde{\alpha}^2}{8m(1+\tilde{\beta}^2)}=-\frac{\alpha^2m}{2\hbar^2(1+\beta^2m^2/\hbar^2)^2}.
\end{eqnarray}
This result coincides with the one of the article \cite{Gadella:2009}.

Having a quantum wave impedance function of a system we can find a wave function of a bound state:
\begin{eqnarray}
\psi(x)=A\exp\left[\frac{im}{\hbar}\int\left(-z_0\theta(-x)+z_0\theta(x)\frac{}{}\right)dx +f0 \right]=\exp\left[\varkappa_0x(\theta(x)-\theta(-x)+1)+f0 \right], 
\end{eqnarray}
where $\varkappa_0=\sqrt{-2mE}/\hbar$ ($E<0$) and
\begin{eqnarray}
f0&=&\lim_{\varepsilon\rightarrow 0}\frac{im}{\hbar}\left\{\theta(-x)\int\limits_{-\varepsilon}^0 Z(x)+\theta(x)\int\limits_{0}^\varepsilon Z(x)\right\}=\ln\left[\psi(0+)\right]\theta(x)+\ln\left[\psi(0-)\right]\theta(-x)=\nonumber\\
&=&\ln\left[1+\tilde{\beta}\right]\theta(x)+\ln\left[1+\tilde{\beta}\right]\theta(-x),
\end{eqnarray}
where $\theta(x)$ is a Heaviside step function.
Finally, we have
\begin{eqnarray}
\psi(x)=A\left\{(1\!-\!\tilde{\beta})\exp[\varkappa_0x]\theta(-x)\!+\!(1\!+\!\tilde{\beta})\exp[-\varkappa_0x]\theta(x)\right\}.
\end{eqnarray}
A normalization condition for a wave function gives the value of $A$ and finally we get
\begin{eqnarray}
\psi(x)\!=\!\frac{\sqrt{\tilde{a}/2}}{1\!+\!\tilde{\beta}^2}\!\left\{\!(1\!-\!\tilde{\beta})\!\exp[\varkappa_0x]\theta(-x)\!+\!(1\!+\!\tilde{\beta})\!\exp[-\varkappa_0x]\theta(x)\!\right\}\!.\nonumber\\
\end{eqnarray} 
This expression also coincides with a well-known result \cite{Gadella:2009}.

\section{Scattering case for a $\delta-\delta'$ potential}

In this case the solution of the equation for a quantum wave impedance function is
\begin{eqnarray}
Z(x)=\left\{\begin{array}{c}
z_0\th[ik_0x+\phi], \ \ x< 0\\
z_0, \ \  x>0
\end{array}\right.,
\end{eqnarray}
where $k_0=\sqrt{2mE}/\hbar$ ($E>0$).
Matching condition (\ref{Z_d_d'_mc}) gives
\begin{eqnarray}
z_0=\frac{i\hbar\tilde{\alpha}}{m(1+\tilde{\beta})^2}+
\frac{(1-\tilde{\beta})^2}
{(1+\tilde{\beta})^2}z_0\th[\phi].
\end{eqnarray}
Using that a wave reflection amplitude coefficient $r=\exp[-2\phi]$ \cite{Arx1:2020} and
\begin{eqnarray}
\th[\phi]=\frac{1-\exp[-2\phi]}{1+\exp[-2\phi]}=\frac{1-r}
{1+r}
\end{eqnarray}
we get
\begin{eqnarray}
r=\frac{4\tilde{\beta}z_0+i\hbar\tilde{\alpha}/m}{2(1+\tilde{\beta}^2)z_0-i\hbar\tilde{\alpha}/m}
\end{eqnarray}
and
\begin{eqnarray}
t=1+r=\frac{2(1+\tilde{\beta})^2z_0}{2(1+\tilde{\beta}^2)z_0-i\hbar\tilde{\alpha}/m}.
\end{eqnarray}
Taking into account the expressions for $\tilde{\alpha}$, $\tilde{\beta}$ and that $z_0=\sqrt{2E/m}$ ($z_0$ is real) we get for a reflection $R$ and a transmission $T$ coefficients the following expressions: 	
\begin{eqnarray}
R=\frac{16\tilde{\beta}^2z_0^2+\hbar^2\tilde{\alpha}^2/m^2}{4(1+\tilde{\beta}^2)^2z_0^2+\hbar^2\tilde{\alpha}^2/m^2}
=\frac{4\tilde{\beta}^2+\Omega^2}{(1+\tilde{\beta}^2)^2+\Omega^2}
,\nonumber\\
T=\frac{4(1+\tilde{\beta})^4z_0^2}{4(1+\tilde{\beta}^2)^2z_0^2+\hbar^2\tilde{\alpha}^2/m^2}=
\frac{(1+\tilde{\beta})^4}{(1+\tilde{\beta}^2)^2+\Omega^2}
,
\end{eqnarray}
where $\Omega= m \alpha/(\hbar^2 k_0)$.

\section{Arbitrary combination of a piecewise constant potential and $\delta$-potentials}
In article \cite{Arx3:2020} the application of a quantum wave impedance method for an arbitrary piesewise constant potential was considered. Here we are going to enrich this model with the presence of $\delta$ potentials. So, 
assume that we have an arbitrary combination of a piecewise constant potential and $\delta$-potentials:
\begin{eqnarray}\label{U_pc_d}
U(x)=U_0\theta(a_0-x)+\sum_{k=1}^{N_1} U_i[\theta(x-a_{k-1})-\theta(x-a_{k})]+
\sum_{j=1}^{N_2}\alpha_j\delta(x-b_j)+U_{N_1+N_2+1}
\theta(x-a_{N_1+1}).\nonumber\\
\end{eqnarray}
Before proceeding we need to sort coordinates $a_k$ and $b_j$ in a descending order and then each of them we rename to $x_i$. After this procedure \mbox{we get}
\begin{eqnarray}
U(x)=U_0\theta(x_0-x)+\sum_{i=1}^{N}\left\{\frac{}{}\right.
U_i[\theta(x-x_{i-1})-\theta(x-x_{i})]
\left.+\alpha_i\delta(x-x_i)+U_{N+1}\theta(x-x_{N+1})
\frac{}{}\right\},
\end{eqnarray}
where $\alpha_i=1$ if $x_i=b_j,\forall j$ and $\alpha_i=0$ if $x_i=a_k, \forall k$. 
Notice, it does not mean that $N$ is equal to $N_1+N_2$ because some $a_k$ can be equal \mbox{to $b_j$.}

The solution of the equation for a quantum wave impedance \cite{Arx1:2020} with the potential (\ref{U_pc_d}) we find in the same form as in \cite{Arx3:2020}].
The substitution of it into the equation for a quantum wave impedance gives
almost the same system of equations as we had in \cite{Arx3:2020} but it differs with the additions $\frac{2i}{\hbar}\alpha_i$ on the right side of the system of equations.
So, finally, have the following relations
\begin{eqnarray}\label{iter_with_d}
-z_m\th\left[\gamma_mx_m+\phi_m\right]+z_{m+1}\th\left[\gamma_{m+1}x_i+\phi_{m+1}\right]=i\frac{2\alpha_m}{\hbar}.
\end{eqnarray}
This system of equations is related to the iterative process of a quantum wave impedance calculation. But now there is discontinuous at the origins of $\delta$-functions. It means that we have to introduce both a left-side $Z(x_m-0)$ and a right-side $Z(x_m+0)$ limits of a quantum wave impedance function at the origins of $\delta$-potentials, namely to the left and to the right sides of points $x=x_m$:
\begin{eqnarray}
&&z_{m}\th\left[\gamma_{m}(x_{m}+c_{m})\right]=Z(x_{m}-0),\nonumber\\
&&z_{m+1}\th\left[\gamma_{m+1}x_{m}+\phi_{m+1}\right]=Z(x_{m}+0).
\end{eqnarray}
Let's write two consecutive equations of the system (\ref{iter_with_d}): 
\begin{eqnarray}\label{iter_con_d}
&&\!\!\!\!\!\!\!\!\!\!\!-z_{m-1}\!\th\!\left[\gamma_{m-1}x_{m-1}\!+\!\phi_{m-1}\right]\!+\!z_m\!\th\left[\gamma_mx_{m-1}\!+\!\phi_m\right]\!=\!i\frac{2\alpha_{m-1}}{\hbar},\nonumber\\
&&\!\!\!\!\!\!\!\!\!\!\!-z_{m}\!\th\!\left[\gamma_{m}x_{m}\!+\!\phi_{m}\right]\!+\!z_{m+1}\!\th\!\left[\gamma_{m+1}x_{m}\!+\!\phi_{m+1}\right]\!=\!i\frac{2\alpha_m}{\hbar}.
\end{eqnarray}
Now our task is to find the relation between $Z(x_{m-1}\pm 0)$, $Z(x_{m}\pm 0)$ and  $\eta_m=i\frac{2\alpha_m}{\hbar}$. 
For this we express $\th\left[\gamma_{m}x_{m}+\phi_{m}\right]$ from the second equation of (\ref{iter_con_d})
and substitute it into the first one. After simple transformations we get:
\begin{eqnarray}
&&Z(x_{m-1}\!-\!0)\!=\!-\!\eta_{m-1}\!+\!
z_m\frac{Z(x_{m}\!+\!0)\!-\!\eta_m\!-\!z_m\th\left[\gamma_{m}\Delta x_m\right]}{z_m\!-\![Z(x_{m}\!+\!0)\!-\!\eta_m]\th\left[\gamma_{m}\Delta x_m\right]}.\nonumber\\
\\
&&Z(x_{m-1}-0)=-\eta_{m-1}+
z_m\frac{Z(x_{m}-0)-z_m\th\left[\gamma_{m}\Delta x_m\right]}{z_m-Z(x_{m}-0)\th\left[\gamma_{m}\Delta x_m\right]}.\nonumber\\
\\
&&Z(x_{m-1}+0)=
z_m\frac{Z(x_{m}-0)-z_m\th\left[\gamma_{m}\Delta x_m\right]}{z_m-Z(x_{m}-0)\th\left[\gamma_{m}\Delta x_m\right]}.\nonumber\\
\\
&&Z(x_{m-1}+0)=
z_m\frac{Z(x_{m}+0)-\eta_m-z_m\th\left[\gamma_{m}\Delta x_m\right]}{z_m-[Z(x_{m}+0)-\eta_m]\th\left[\gamma_{m}\Delta x_m\right]}.\nonumber
\\
\end{eqnarray}
We can write these four relations as one in the following way
\begin{eqnarray}\label{Z_calc_r_to_l}
Z(x_{m-1}\pm_10)=-\eta_{m-1}(1\mp_11)/2+z_m\frac{Z(x_{m}\pm_20)-\eta_m(1\pm_21)/2-z_m\th\left[\gamma_{m}\Delta x_m\right]}{z_m-[Z(x_{m}+0)-\eta_m]\th\left[\gamma_{m}\Delta x_m\right]}.
\end{eqnarray}
In a case of a wave propagating in the opposite direction, namely, in the negative direction of $x$ axis we get
\begin{eqnarray}\label{Z_calc_l_to_r}
Z(x_{m}\pm_2 0)=\eta_{m}(1\pm_21)/2+z_m\frac{Z(x_{m-1}\pm_10)+\eta_{m-1}(1\mp_11)/2+z_m\th\left[\gamma_{m}\Delta x_m\right]}{z_m+[Z(x_{m-1}-0)+\eta_{m-1}]\th\left[\gamma_{m}\Delta x_m\right]},
\end{eqnarray}
where 
\begin{eqnarray}
\Delta x_m=x_m-x_{m-1}.
\end{eqnarray}
Notice, that if $\eta_m=0$ then $Z(x_m-0)=Z(x_m+0)$ since $Z(x_m+0)=Z(x_m-0)+\eta_m$.

\section{Conclusions}

Zero-range singular-potentials, in particular $\delta$-potential, are used very often for modelling  periodic structures and different inhomogeneities. It is due to the fact the models which are based on a $\delta$-potential and a piecewise constant potential allow obtaining exact analytical solutions (for a wide range of model systems) and investigating important features of the studied systems (in particular crystal-like structures). By using a $\delta$-potential we can model a zone structure of crystal lattices, a density of its states, an influence of impurities and edges of crystals on its spectrum etc. For example, a periodic lattice of $\delta$ inhomo\-geneities is a convenient model for
analysing the characteristics of crystals and crystal-like structures. In \cite{Nelin_Liashok:2015} $\delta$-inhomogeneities are used for modeling micro- and nanostructures and the criteria of impedance inhomogeneities approaching by $\delta$-inhomo\-geneities are established. By the way, it was shown that $\delta$-
inhomogeneities introduce a reactive component of a quantum wave impedance.

In this area of research a quantum wave impedance approach demonstarted its efficacy  \cite{Vodolazka_Nelin:2013, Nelin_Liashok:2015, Gindikina_Zinger_Nelin:2015,Nelin_Shulha_Zinher:2017, Nelin_Liashok:2016, Nelin_Shulha_Zinher:2018, Babushkin_Nelin:2011}. But untill this paper there were no a consistent theory which allow including into the consideretation diferent types of zero-range singular potential as well as formalizing the process of solving quantum-mechanical systems with arbitrary combination of a piesewise constant potential and singular zero-range potentials. The results of this paper can be effectively applied for the numerical calculations of physical parameters of the nano-sctructures with complicated geometry of potential energy within the approach of a quantum wave impedance.

\renewcommand\baselinestretch{1.0}\selectfont


\def\name{\vspace*{-0cm}\LARGE 
	Bibliography\thispagestyle{empty}}
\addcontentsline{toc}{chapter}{Bibliography}

{\small

	\bibliographystyle{gost780u}
	\bibliography{full.bib}
	
}

\newpage

\end{document}